\documentclass[5p,authoryear,twocolumn]{elsarticle}

\usepackage{journals}
\usepackage{natbib}
\usepackage[colorlinks=true,linkcolor=red,urlcolor=blue,citecolor=gray]{hyperref}
\usepackage{algorithm,algorithmic}
\usepackage{upgreek}

\begin{document}

\title{Enabling pulsar and fast transient searches using coherent
  dedispersion}

\author[astron]{C.\,G.\,Bassa}
\ead{bassa@astron.nl}
\author[api]{Z.\,Pleunis}
\author[astron,api]{J.\,W.\,T.\,Hessels}

\address[astron]{ASTRON, the Netherlands Institute for Radio
  Astronomy, Postbus 2, 7990 AA, Dwingeloo, The Netherlands}
\address[api]{Anton Pannekoek Institute for Astronomy, University of
  Amsterdam, Science Park 904, 1098 XH, Amsterdam, The Netherlands}

\begin{abstract}
  We present an implementation of the coherent dedispersion algorithm
  capable of dedispersing high-time-resolution radio observations to
  many different dispersion measures (DMs). This approach allows the
  removal of the dispersive effects of the interstellar medium and
  enables searches for pulsed emission from pulsars and other
  millisecond-duration transients at low observing frequencies and/or
  high DMs where time broadening of the signal due to dispersive
  smearing would otherwise severely reduce the sensitivity.  The
  implementation, called \texttt{cdmt}, for \textit{coherent
    dispersion measure trials}, exploits the parallel processing
  capability of general-purpose graphics processing units to
  accelerate the computations. We describe the coherent dedispersion
  algorithm and detail how \texttt{cdmt} implements the algorithm to
  efficiently compute many coherent DM trials. We present the concept
  of a semi-coherent dedispersion search, where coherently dedispersed
  trials at coarsely separated DMs are subsequently incoherently
  dedispersed at finer steps in DM. The software is used in an ongoing
  LOFAR pilot survey to test the feasibility of performing
  semi-coherent dedispersion searches for millisecond pulsars at
  135\,MHz. This pilot survey has led to the discovery of a radio
  millisecond pulsar -- the first at these low frequencies. This is
  the first time that such a broad and comprehensive search in
  DM-space has been done using coherent dedispersion, and we argue
  that future low-frequency pulsar searches using this approach are
  both scientifically compelling and feasible. Finally, we compare the
  performance of \texttt{cdmt} with other available alternatives.
\end{abstract}


\maketitle

\section{Introduction}
Advances in electronics, computing and networking, primarily following
Moore's law, have enabled the realization of a new type of radio
telescope. Instead of placing receivers at the focus of movable
reflecting dishes, these new telescopes employ a large number of
stationary dipole antennas to create what is called an
\textit{aperture array}. The signals from these antennas are combined
digitally in a correlator to create images of the sky, or in a
beamformer to form beams on the sky. Three major digital aperture
arrays operating at low radio frequencies are presently in operation:
the LOw Frequency Array (LOFAR; \citealt{hwg+13}), the Murchison Wide
Field Array (MWA; \citealt{lcm+09,tgb+13}) and the Long Wavelength
Array (LWA; \citealt{ecc+09,etc+13}), while the low-frequency
component of the Square Kilometre Array (SKA1-Low) is being planned
\citep{bbg+15}. To maximize sensitivity while keeping the number of
individual antennas, and hence cost, low, these aperture arrays
operate at long wavelengths, and hence low observing frequencies
(below 300\,MHz).

One of the key science areas at these low observing frequencies is the
study of radio pulsars. Radio emission from these highly magnetized,
rotating neutron stars exhibits a very steep spectrum
\citep{mkkw00,blv13}, typically peaking or turning over between 100 to
200\,MHz (e.g.\,\citealt{mgj+94}). Surveys at these low frequencies
have the prospect of discovering new pulsars that are too faint to be
detected in surveys performed at higher observing frequencies and can
take advantage of large fields of view (e.g.\,\citealt{clh+14}). Of
particular interest are radio pulsars spinning at millisecond spin
periods. These millisecond pulsars provide unparalleled
precision for measuring neutron star masses, performing precision
tests of General Relativity, understanding binary evolution, and
detecting gravitational waves
(e.g.\ \citealt{ksm+06,dpr+10,rsa+14,afw+13,vlh+16}).

Radio emission propagating through the ionized interstellar medium
suffers from dispersion, introducing a frequency dependent time delay
over the requisite large bandwidths of radio astronomical
observations. As a result, pulsed signals, such as those of pulsars
and fast transients (a generic term used for other sources of
millisecond-duration radio pulses; e.g.\,\citealt{lbm+07}), have a
specific dispersion measure ($\mathrm{DM}$) which relates directly to
the column density of free electrons between the source and the
observer. When searching for new pulsars and fast transients,
correcting for this dispersion is typically done through
\textit{incoherent dedispersion}, where the dispersive delays are
removed by time shifting the time-series of individual, narrow,
frequency channels by an amount appropriate to the DM of the
source. Though this corrects for dispersion between channels, the
dispersion within the finite bandwidth of the individual channels is
not corrected for. Nonetheless, the computational efficiency of the
technique has been a practical necessity compared to more accurate
approaches.

A priori, the DM of a new pulsar or fast transient is unknown, and the
data must be dedispersed to a broad range of different DMs. Models for
the Galactic electron density (e.g.\,\citealt{cl02}) can be used to
estimate the maximum DM towards a given direction though -- to enable
sensitivity to extra-Galactic fast transients -- most ongoing surveys
search up to a maximum DM of several thousand pc\,cm$^{-3}$. Depending
on the frequency and time resolution of the input data, several
thousand DM trials may need to be computed \citep{cm03}\footnote{See
  \citealt{lsc+16b} and
  \url{http://www.jb.man.ac.uk/pulsar/Surveys.html} for the parameters
  of ongoing and historic pulsar and fast transient surveys.}. Though
dedispersing that many DM trials is a computationally expensive task,
recent implementations of incoherent dedispersion algorithms on
graphics processing units (GPUs) are fast enough to allow real-time
processing \citep{mks+11,bbbf12,slbn16}.

\begin{figure}
  \includegraphics[angle=270,width=\columnwidth]{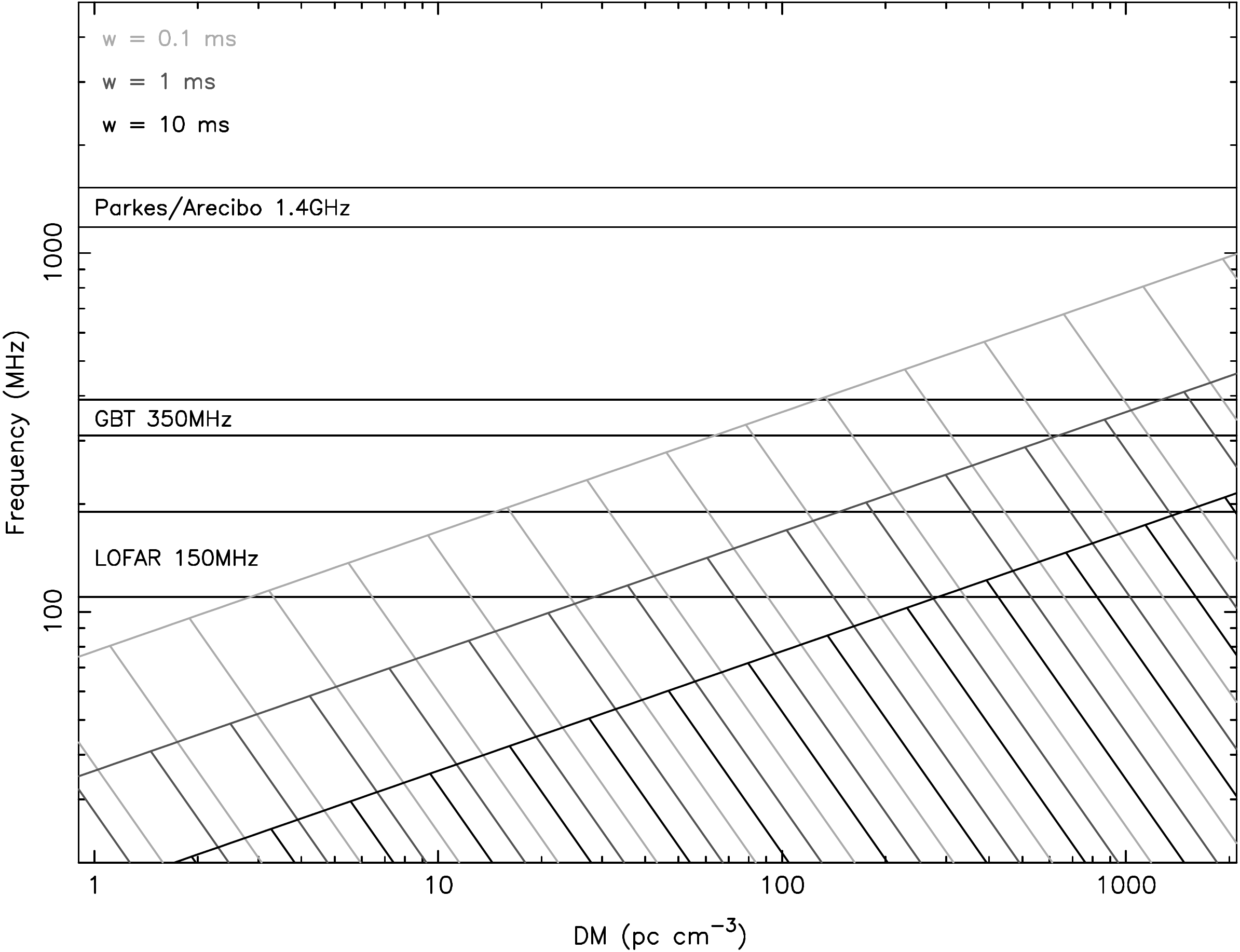}
  \caption{The effect of residual dispersion smearing in frequency
    channels as a function of dispersion measure (DM) and
    frequency. The diagonal lines denote the $3\sigma$ detection limit
    of an undispersed input pulse of $10\sigma$ with pulse
    full-width-half-maxima of $w=0.1$, 1 and 10\,ms. For frequencies
    below, or DMs above, these limits, denoted by the hashed areas,
    the pulse is no longer detectable. The channel size is set at
    $\Delta \nu=0.02$\,MHz which corresponds to a time resolution of
    $\Delta t=\Delta \nu^{-1}=50$\,$\upmu$s. This time resolution is
    what is typically used in current millisecond pulsar
    searches. Using narrower channels would adversely reduce the time
    resolution. The frequency bands and central frequencies of
    representative radio telescopes are shown with the horizontal
    lines.}
  \label{fig:dmf}
\end{figure}

At low frequencies and/or high DMs, incoherent dedispersion can lead
to significant smearing of the pulse (in time) within a channel (see
Fig.\,\ref{fig:dmf}). The effect of dispersion can be completely
removed though \textit{coherent dedispersion}. This approach convolves
the input signals with the inverse of the transfer function of the
interstellar medium. This convolution must be performed before the
signal is detected (squared) as the phase information, in addition to
the amplitude, is required. Hence, the data rate and computational
requirements for coherent dedispersion are typically larger than the
filterbanked data used for incoherent dedispersion, as for the latter
the two polarizations can be squared and frequency and/or time
resolution can be reduced. Because of these higher data rates,
coherent dedispersion is presently only used for observing either
known pulsars or when searching for pulsars in globular clusters with
known DMs.

Here we present \texttt{cdmt}, for \textit{coherent dispersion measure
  trials}, which implements the coherent dedispersion algorithm to
perform coherent dedispersion to many dispersion measure trials in
parallel on GPUs. This software allows us to control the residual
dispersion smearing within a channel and retain both high time and
high frequency resolution when searching for pulsars and fast
transients. In a semi-coherent dedispersion search, the input data can
be coherently dedispersed to several coarsely separated trial DMs,
each of which is then incoherently dedispersed with finer DM steps
around the coherent trial DM. Though the total number of incoherent DM
trials, and hence processing requirements, will increase, this
approach allows us to search for millisecond pulsars at lower
observing frequencies than were previously possible -- thus probing a
new astrophysical parameter space.

The paper is structured as follows. The coherent dedispersion
algorithm, combined with channelizing the data, is described in
\S\,\ref{sec:description}. Our implementation of the algorithm is
outlined in \S\,\ref{sec:implementation}. In \S\,\ref{sec:application}
we provide an application example of the software and we report on the
performance in \S\,\ref{sec:performance}.  Finally, we discuss
prospects for \texttt{cdmt} in \S\,\ref{sec:discussion}.

\section{Algorithm description}\label{sec:description}
\begin{figure}
  \includegraphics[angle=270,width=\columnwidth]{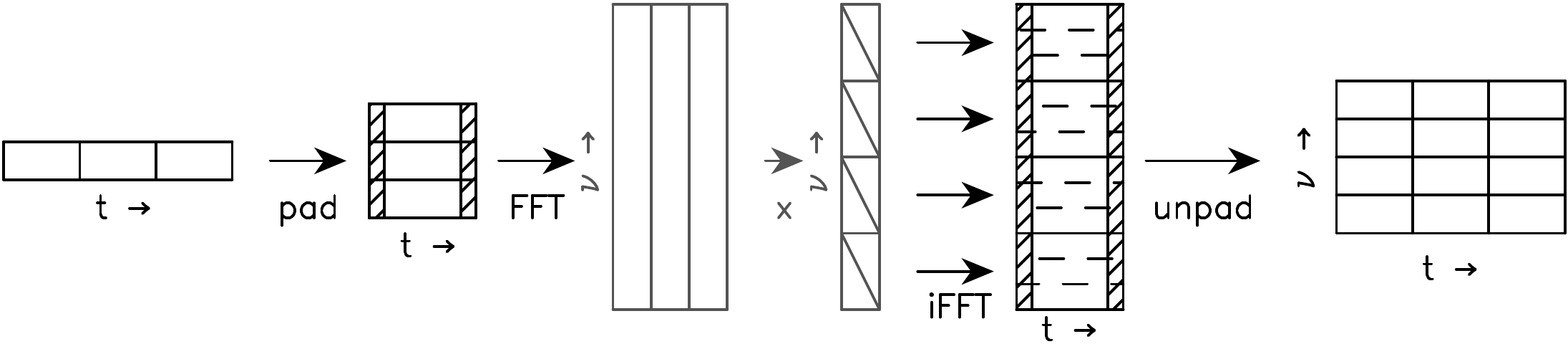}
  \caption{A schematic diagram depicting coherent dedispersion
    in a convolving synthetic filterbank. From left to right, the
    Nyquist sampled input time-series is broken up into sections of
    length $N_\mathrm{bin}$ (stage 1), which are overlapped by
    $n_\mathrm{d}n_\mathrm{c}/2$ time-samples at either end (stage
    2). These sections are then Fourier transformed to the frequency
    domain (stage 3), and multiplied by the frequency domain transfer
    function of $n_\mathrm{c}$ channels (stage 4). A Fourier transform
    of length $M_\mathrm{bin}=N_\mathrm{bin}/n_\mathrm{c}$ back to the
    time domain is then performed for each channel to transform back
    to the time domain (stage 5). The overlap region of $n_\mathrm{d}$
    time samples of each channelized time-series is then discarded
    (unpadded) to form the coherently dedispersed filterbank (stage
    6). The black stages are in the time domain, the frequency domain
    stages are colored in gray.}
  \label{fig:cartoon}
\end{figure}

The coherent dedispersion algorithm as implemented in \texttt{cdmt} is
that of a convolving synthetic filterbank. This implementation
performs coherent dedispersion as a complex multiplication in the
frequency domain and combines dedispersion with channelization. For a
detailed description of the coherent dedispersion algorithm and
different filterbanking versions, we refer to the implementation in
the \texttt{dspsr} software package, which is detailed in
\citet{sb10}\footnote{See also Willem van Straten's PhD thesis at
  \url{
 http://astronomy.swin.edu.au/~wvanstra/papers/thesis.html}}. Here, we
closely follow the description outlined in these references to explain
our implementation. A schematic representation of the implementation
is depicted in Figure\,\ref{fig:cartoon}.

Coherent dedispersion is a convolution of the raw signal voltages
(Nyquist sampled time-series) with the inverse of the transfer
function of the interstellar medium (ISM). The convolution is most
efficiently performed as a multiplication in the frequency domain
through the discrete convolution theorem
\citep[see][Chapter\,13.1]{ptvf92}. In addition to the convolution,
the convolving synthetic filterbank trades time resolution for
frequency resolution to create a user-defined number of channels
$n_\mathrm{c}$ over the input bandwidth. This channelization step is
combined with coherent dedispersion by performing a large forward
Fourier transform of $N_\mathrm{bin}$ samples prior to dedispersion,
followed by $n_\mathrm{c}$ backward Fourier transforms of
$N_\mathrm{bin}/n_\mathrm{c}$ samples to provide the channelization.

The transfer function of the ISM, when modelled as a cold tenuous
plasma, is defined in the frequency domain (\citealt{han71,hr75}, see
also \citealt{lk12}) as:
\begin{equation}\label{eqn:coherent}
H(\nu+\nu_0)=\exp \left [\frac{2\pi i \nu^2 k_\mathrm{DM}\mathrm{DM}}{\nu_0^2
    (\nu+\nu_0)} \right ].
\end{equation}
Here $\nu_0$ is the center frequency of a subband or channel of
bandwidth $\Delta \nu$ (both in MHz), while $\nu$ is the frequency
offset within the channel or subband, such that $-\Delta
\nu/2<\nu<\Delta \nu/2$.  The dispersion measure is $\mathrm{DM}$ (in
pc\,cm$^{-3}$) and $k_\mathrm{DM}^{-1}=2.41\times10^{-4}$\,s\,MHz$^2$
is a measured constant of proportionality \citep{mt72}. The duration
of the transfer function is determined by the dispersion sweep
$t_\mathrm{d}$ for the DM and parameters of the subband or channel
($\nu_0$, $\Delta \nu$) being dedispersed,
\begin{equation}\label{eqn:incoherent}
  t_\mathrm{d}=k_\mathrm{DM} (\nu_\mathrm{min}^{-2}-\nu_\mathrm{max}^{-2})\mathrm{DM}.
\end{equation}
Here $\nu_\mathrm{min}=\nu_0-\Delta\nu/2$ and
$\nu_\mathrm{max}=\nu_0+\Delta\nu/2$. In \texttt{cdmt}, an input
subband is channelized to $n_\mathrm{c}$ channels, each of which is
coherently dedispersed with a transfer function
(Eqn.\,\ref{eqn:coherent}) appropriate for that channel.

Owing to the underlying assumptions of the discrete convolution
theorem, the edges of the output time-series will be polluted by the
wrap around region. To prevent this, we use the \textit{overlap-save}
method, which requires each section of the time-series (raw signal
voltages) of an input subband to be overlapped by at least
$n_\mathrm{d} n_\mathrm{c}$ time samples \citep[see][Fig.\,2 and
  3]{sb10}. A total of $n_\mathrm{d}$ channelized time samples are
subsequently discarded from the time-series of each output channel.

Given the sample rate $r$ of the input subband, which is channelized
to $n_\mathrm{c}$ output channels, the size of the overlap region is
determined by $n_\mathrm{d}=r t_\mathrm{d}/n_\mathrm{c}$. Here the
dispersion sweep $t_\mathrm{d}$ is for the channel with the lowest
frequency. The size $N_\mathrm{bin}$ of the forward Fourier transform
can now be chosen such that $N_\mathrm{bin}>n_\mathrm{d}n_\mathrm{c}$,
while the $n_\mathrm{c}$ backward Fourier transforms have size
$M_\mathrm{bin}=N_\mathrm{bin}/n_\mathrm{c}$.

\section{Implementation}\label{sec:implementation}
Our aim was to develop software capable of efficiently computing
synthetic filterbanks that are coherently dedispersed to many
different DMs. To our knowledge, no such software currently exists. To
accelerate the computations, we chose to perform the majority of the
computations on many-core GPUs and develop our software using the
NVIDIA Compute Unified Device Architecture
(CUDA)\footnote{\url{http://www.nvidia.com/object/cuda_home_new.html}}
programming model.  CUDA provides the \textsc{cuFFT}
library\footnote{\url{http://developer.nvidia.com/cufft}}, which
contains algorithms for performing large numbers of Fast Fourier
Transforms (FFTs) in parallel on GPUs. Further acceleration was
achieved by performing the majority of the computations in kernels
that can be executed in parallel on the GPU. Intermediate results,
such as reading the input data and performing the forward FFT, were
reused to further boost efficiency. The implementation of the
algorithm is given in Algorithm\,\ref{alg:implementation}.

In the following, we will consider the input data to consist of
$n_\mathrm{s}$ dual-polarization, complex-valued, Nyquist-sampled
subbands. Each of these subbands will be channelized to $n_\mathrm{c}$
channels.

\begin{algorithm}
  \caption{\texttt{cdmt} pseudo algorithm}\label{alg:implementation}
  \begin{algorithmic}
    \STATE initialize (read metadata)
    \FOR{idm = 0 $\rightarrow$ ndm }
    \STATE compute transfer function for DM[idm]
    \ENDFOR

    \FOR{iblock = 0 $\rightarrow$ nblock }
    \STATE read block
    \STATE copy to device
    \STATE unpack, segment and pad
    \STATE forward FFT
    \STATE swap halves per subband
    \FOR{idm = 0 $\rightarrow$ ndm}
    \STATE complex multiplication with transfer function for DM[idm]
    \STATE swap halves per channel
    \STATE backward FFT
    \STATE transpose, unpad and detect
    \STATE compute block sums
    \STATE compute channel statistics
    \STATE redigitize
    \STATE copy to host
    \STATE write to disk
    \ENDFOR
    \ENDFOR
  \end{algorithmic}
\end{algorithm}

Initial processing steps consist of reading the metadata of the
$n_\mathrm{s}$ input subbands, as well as initializing user-defined
choices for the number of channels $n_\mathrm{c}$ per subband, the
forward FFT size $N_\mathrm{bin}$, and the size of the overlap region
$n_\mathrm{d}$ appropriate for the input data. We have specifically
chosen to keep $N_\mathrm{bin}$ and $n_\mathrm{d}$ identical for all
subbands and channels. Though this reduces the efficiency of the
algorithm, as more unpolluted time samples will be discarded, it also
greatly reduces the complexity of the implementation, and allows
multiple channels and subbands to be processed in parallel. Given the
choices for $N_\mathrm{bin}$, $n_\mathrm{c}$ and $n_\mathrm{d}$ and
the frequency setup of the input subbands, the transfer functions for
each channel and each of the $n_\mathrm{DM}$ coherent DM trials are
computed on the GPU.

The program then enters a loop where the input time-series of the
subbands are read in blocks of
$N_\mathrm{FFT}(N_\mathrm{bin}-n_\mathrm{d}n_\mathrm{c})$ time-samples
at a time, which are copied to the GPU. The value for
$N_\mathrm{FFT}$, the number of FFTs performed in parallel, is chosen
to maximize the usage of the GPU memory. Once on the GPU, the input
data blocks are unpacked into two \textsc{cufftComplex} arrays, one
for each polarization. These data blocks are then segmented and padded
to $N_\mathrm{bin}$ samples with an overlap region of
$n_\mathrm{d}n_\mathrm{c}/2$ samples at each edge, forming a cube of
$N_\mathrm{bin} N_\mathrm{FFT} n_\mathrm{s}$ samples for each
polarization. These operations are performed in a single GPU kernel,
where each axis of the cube maps directly onto the 3 dimensions of a
grid of GPU thread blocks.

The \textsc{cuFFT} library is then used to perform two in-place, batch
complex-to-complex forward FFTs of size $N_\mathrm{bin}$, resulting in
$N_\mathrm{FFT} n_\mathrm{s}$ spectra of $N_\mathrm{bin}$
complex frequency bins each, one for each polarization. Since the
output of the complex-to-complex FFT places the positive frequencies
in the first half of the array and the negative frequencies in the
second half, a GPU kernel swaps these halves around to arrange the
frequency bins in increasing frequency.

Next, the program enters a second loop, this time over the
$N_\mathrm{DM}$ different DM trials. In each iteration of this loop,
the $N_\mathrm{FFT} n_\mathrm{s}$ spectra of $N_\mathrm{bin}$
frequency bins for both polarizations are multiplied by the transfer
functions of the $n_\mathrm{c}$ channels at the DM of that
iteration. Since the transfer functions for different channels are
consecutive, the complex multiplication creates $N_\mathrm{FFT}
n_\mathrm{s} n_\mathrm{c}$ complex spectra of
$M_\mathrm{bin}=N_\mathrm{bin}/n_\mathrm{c}$ frequency bins.

Before performing the backward FFT, the two halves of each spectrum of
$M_\mathrm{bin}$ frequency bins are swapped back, using the same GPU
kernel as before. The \textsc{cuFFT} library is again used to perform
$N_\mathrm{FFT} n_\mathrm{s} n_\mathrm{c}$ complex-to-complex backward
FFTs of size $M_\mathrm{bin}$ in batch mode for both
polarizations. This transforms the coherently dedispersed frequency
bins back to the time domain.

The FFT output is now out of order, with the array indexes moving over
time (data blocks), frequency (subbands), time (samples), and
frequency (channels). A GPU kernel transposes the output to frequency
(subbands and channels) to time (samples and data blocks). This
transpose is combined with three further steps. First, it removes the
effects of the padding step, where for each of the $n_\mathrm{c}$
channels of the $n_\mathrm{s}$ subbands for $N_\mathrm{FFT}$ blocks,
$n_\mathrm{d}/2$ time samples at each edge of the time-series are
discarded. Secondly, the complex valued time-series of the two
polarizations are squared and added to form Stokes I
(intensity). Finally, the Stokes I output is packed such that spectra
are in decreasing frequency order, forming a filterbank of
$N_\mathrm{FFT}(M-n_\mathrm{d})$ spectra of $n_\mathrm{s}
n_\mathrm{c}$ channels.

At this stage it is beneficial to redigitize the floating point output
to 8-bit integers to reduce data rates and disk storage. Data offsets
and scales are computed from the average and standard deviation of the
floating point time-series of each channel of
$N_\mathrm{FFT}(M_\mathrm{bin}-n_\mathrm{d})$ samples. To efficiently
compute these offsets and scales for each channel, a GPU kernel first
computes sums of intensity and intensity squared for segments of $M$
samples in parallel. Next, another GPU kernel uses the sums for these
segments to compute the average and standard deviation of each
channel. A third GPU kernel then uses these values to determine the
offset and scale of each channel, and use them to convert the floating
point values to 8-bit integers. The
$N_\mathrm{FFT}(M_\mathrm{bin}-n_\mathrm{d})n_\mathrm{s}n_\mathrm{c}$
8\,bit values for this data block and this coherent DM trial are
copied back to the CPU and written to disk. The program then continues
the loop over the coherent DM trials, and the loop over the data
blocks until the entire input file has been processed.

\section{Application}\label{sec:application}
The \texttt{cdmt} software was specifically designed for a survey with
the LOFAR radio telescope \citep{sha+11,hwg+13} for millisecond
pulsars associated with unidentified \textit{Fermi} $\gamma$-ray
sources from the 3FGL catalog \citep{aaa+15_3fgl}. Here, we present
results using the data from this survey as input; the full scientific
results of the survey will be presented in a forthcoming paper
(Pleunis et al., in prep.).

The LOFAR survey targeted 52 so-called `unidentified' \textit{Fermi}
$\gamma$-ray sources with the LOFAR HBAs (high band antennas). These
are sources of $\gamma$-ray emission, identified with the
\textit{Fermi Gamma-ray Space Telescope}, for which the astrophysical
origin of the emission is still unclear. Pulsars are one of the prime
candidates for creating such emission, and radio pulsation searches
towards these $\gamma$-ray sources can provide a definitive
identification, as well as the basis for further scientific insights.

Each source was observed for 20\,min, and the LOFAR \textsc{Cobalt}
correlator formed 7 tied-array beams\footnote{A tied-array beam can be
  considered as a field of view on the sky obtained by coherently
  adding the signals from individual LOFAR stations using appropriate
  time and phase delays.} to cover the error regions of the
$\gamma$-ray sources. For each of these tied-array beams, 200
dual-polarization, complex valued, Nyquist sampled subbands of
195.31\,kHz (39.06\,MHz total) bandwidth, centered at a frequency
of 135\,MHz, were recorded to disk. The observations were processed on
the \textsc{Dragnet} GPU cluster, which consists of 23 nodes, each of
which has 4 NVIDIA Titan X GPUs, 128\,GB of RAM and dual 8-core Xeon
E5-2630 CPUs. The nodes are interconnected with 1\,Gb and 10\,Gb
ethernet networks. The observations were stored on 20 of the 23 nodes,
with the bandwidth of each source distributed over 20 parts of 10
consecutive subbands each, with one part stored on each node.

\begin{figure}
  \includegraphics[angle=270,width=\columnwidth]{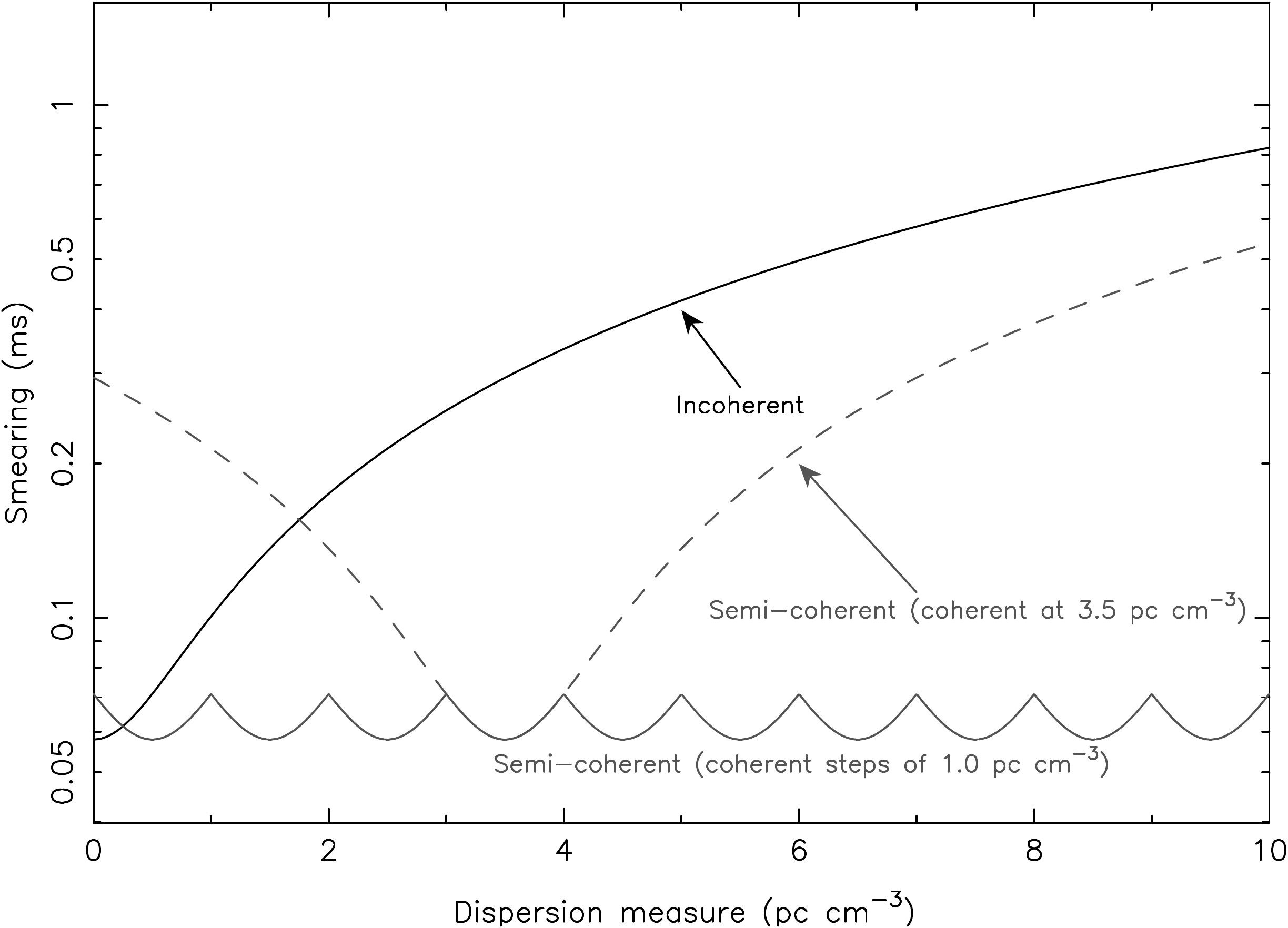}
  \caption{Smearing due to the combined effects of dispersion within a
    channel, finite time resolution, and finite DM steps over the full
    bandwidth, as a function of DM. The survey parameters are as
    described in the text. A fully incoherent dedispersion approach is
    shown with the solid black line. The smearing quickly exceeds
    0.1\,ms for DMs larger than 1\,pc\,cm$^{-3}$. Using a
    semi-coherent dedispersion approach, first coherently dedispersing
    the data to a DM of 3.5\,pc\,cm$^{-3}$ and then using incoherent
    dedispersion, results in the dashed gray line, which is a
    horizontally shifted copy of the black line. Here, the smearing is
    below 0.1\,ms between a DM of 2.5 and 4.5\,pc\,cm$^{-3}$. Using
    coherently dedispersed data at steps of 1\,pc\,cm$^{-3}$, and then
    incoherently dedispersing in between, results in the solid gray
    line; limiting the dispersive smearing to less than 0.07\,ms.}
  \label{fig:ddplan}
\end{figure}

Since the survey specifically targets sources at high Galactic
latitudes, the maximum DM was chosen at 80\,pc\,cm$^{-3}$, which is
approximately twice the maximum DM predicted by the NE2001 Galactic
electron density model \citep{cl02} towards the selected
\textit{Fermi} $\gamma$-ray sources. The 195.31-kHz subbands were
channelized to $n_\mathrm{c}=8$ channels of 24.41\,kHz, yielding a
time resolution of 40.96\,$\upmu$s. No further averaging in frequency
or time was performed. We chose to limit the effects of dispersive
smearing to less than 0.07\,ms, and to allow at least 10 independent
pulse phase bins for a millisecond pulsar with a spin period of
1\,ms. To achieve this, we used \texttt{cdmt} to coherently dedisperse
these observations over 80 coherent DM trials. These DM trials were
spaced at 1\,pc\,cm$^{-3}$ from 0.5\,pc\,cm$^{-3}$ to
79.5\,pc\,cm$^{-3}$. Figure\,\ref{fig:ddplan} shows the dispersive
smearing for the observational setup and semi-coherent dedispersion
approach. The size of the forward FFT was chosen at
$N_\mathrm{bin}=65536$ samples, with an overlap region of
$n_\mathrm{d}=2048$ samples (with $n_\mathrm{c}=8$,
$M_\mathrm{bin}=8192$ samples). The available memory on the NVIDIA
Titan X GPU allowed $N_\mathrm{FFT}=100$ FFTs of $N_\mathrm{bin}$
samples to be coherently dedispersed in parallel.

\begin{figure}
  \includegraphics[angle=270,width=0.49\columnwidth]{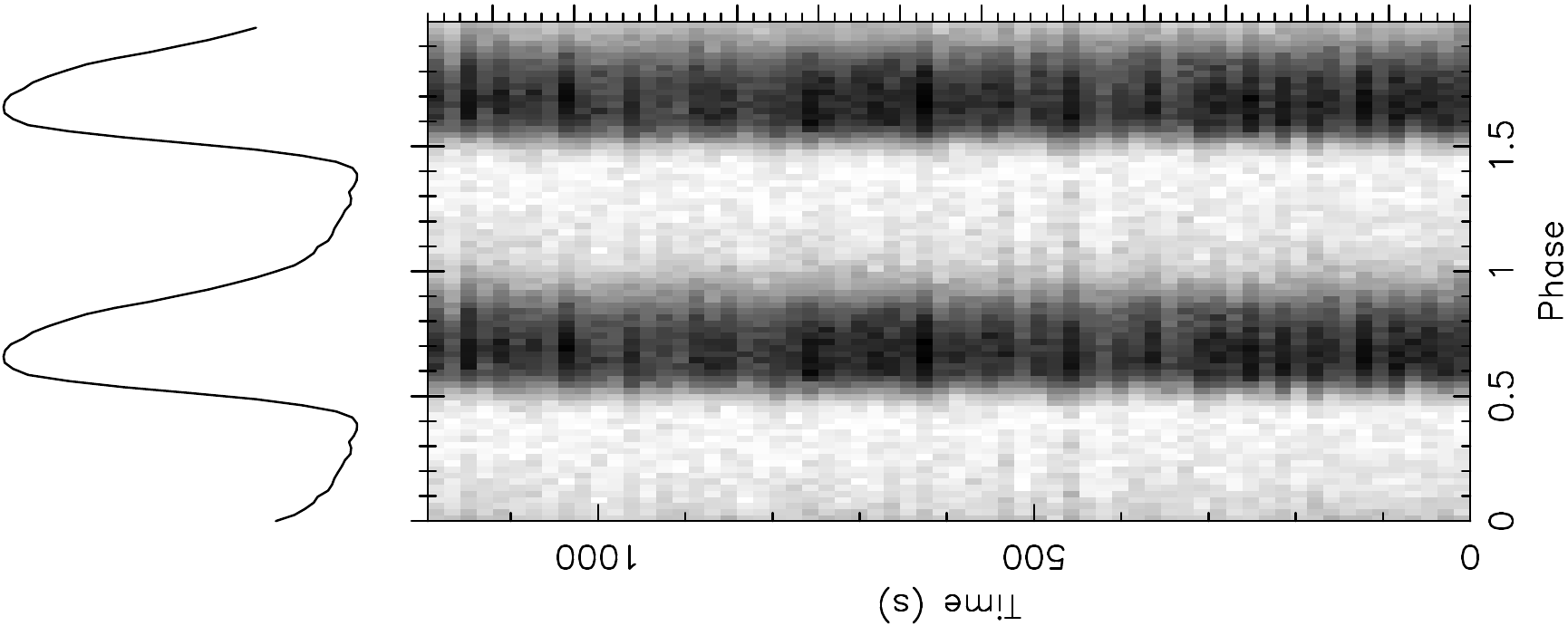}
  \includegraphics[angle=270,width=0.49\columnwidth]{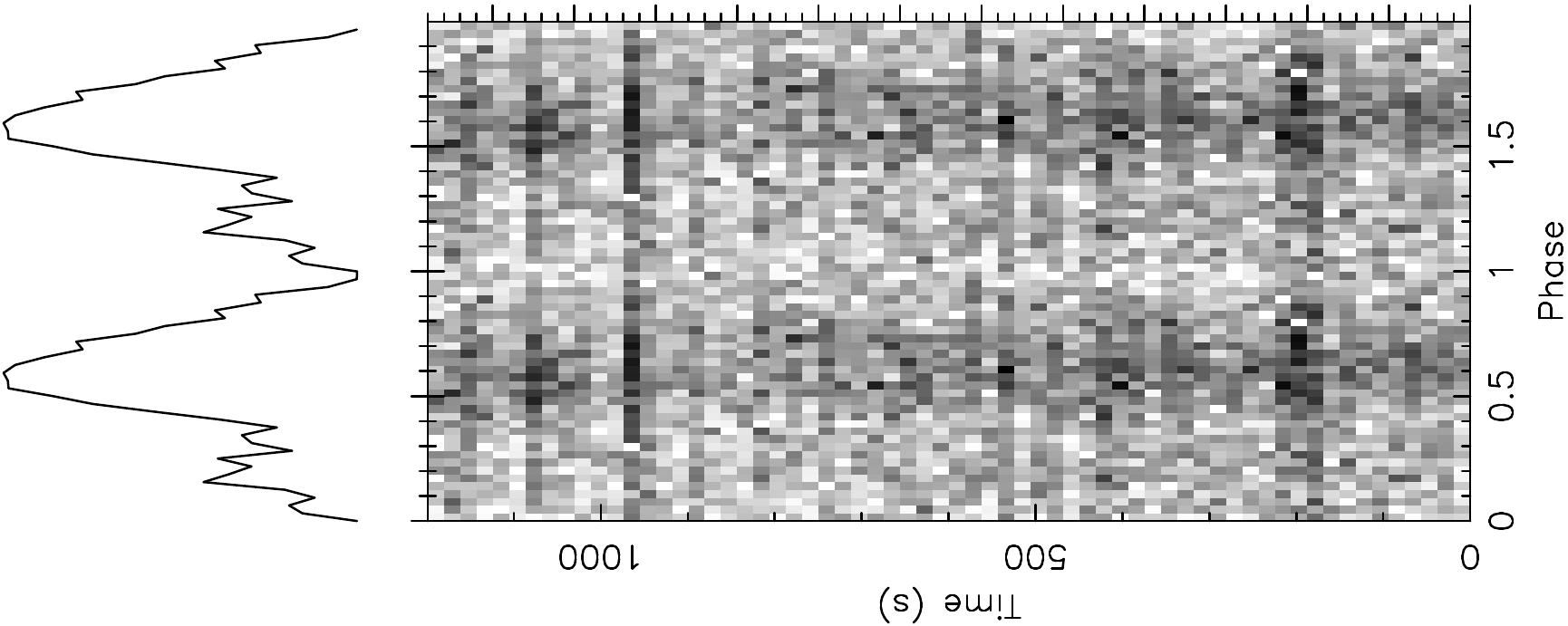}
  \caption{Integrated pulse profiles (top curves; two full cycles are
    shown for clarity) and pulse brightness as a function of time
    (bottom panel) for test observations of PSR\,J1810+1744 (left) and
    PSR\,J2215+5135 (right). These are 20-min observations with the
    LOFAR HBA of 39\,MHz of bandwidth centered
    on 135\,MHz. PSR\,J1810+1744 has $P=1.66$\,ms and
    $\mathrm{DM}=39.658$\,pc\,cm$^{-3}$ and PSR\,J2215+5135 has
    $P=2.61$\,ms and $\mathrm{DM}=69.194$\,pc\,cm$^{-3}$. Both pulsars
    were blindly redetected using the semi-coherent dedispersion pipeline
    presented here.}
  \label{fig:msps}
\end{figure}

\begin{figure}
  \includegraphics[angle=270,width=0.49\columnwidth]{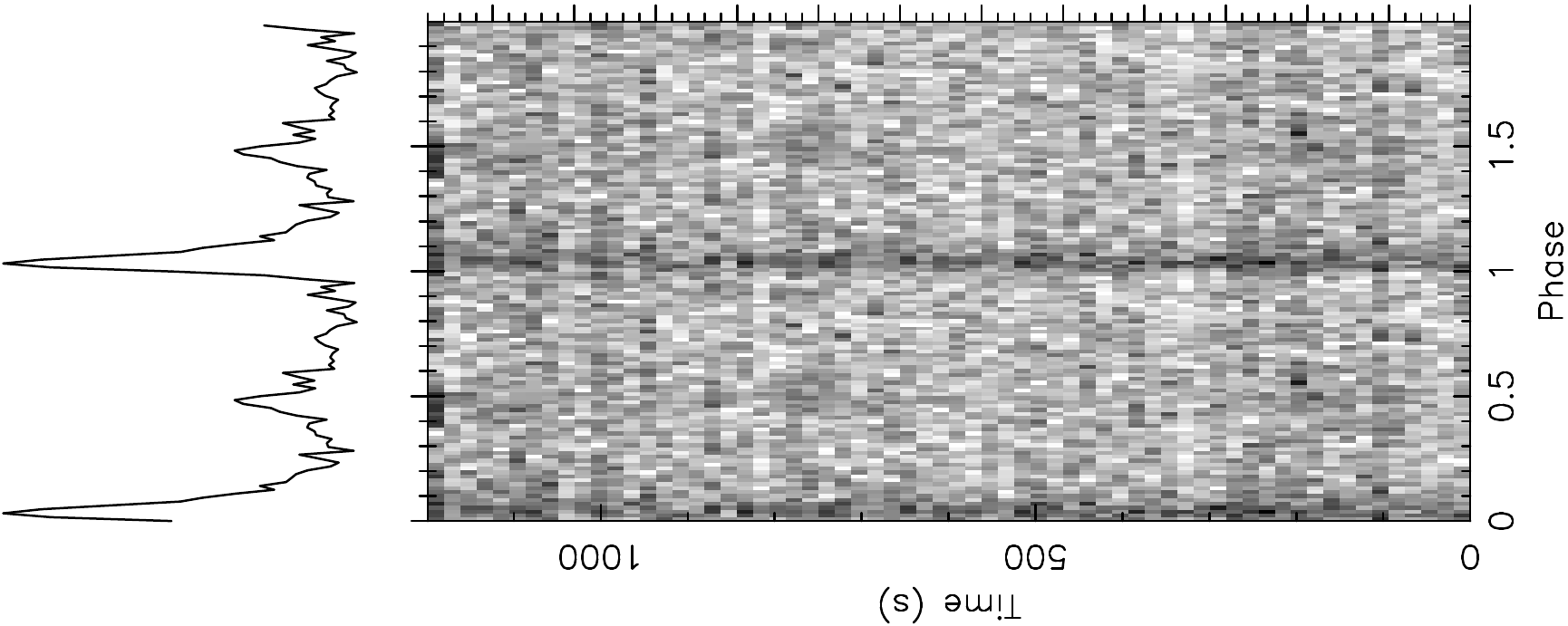}
  \includegraphics[angle=270,width=0.49\columnwidth]{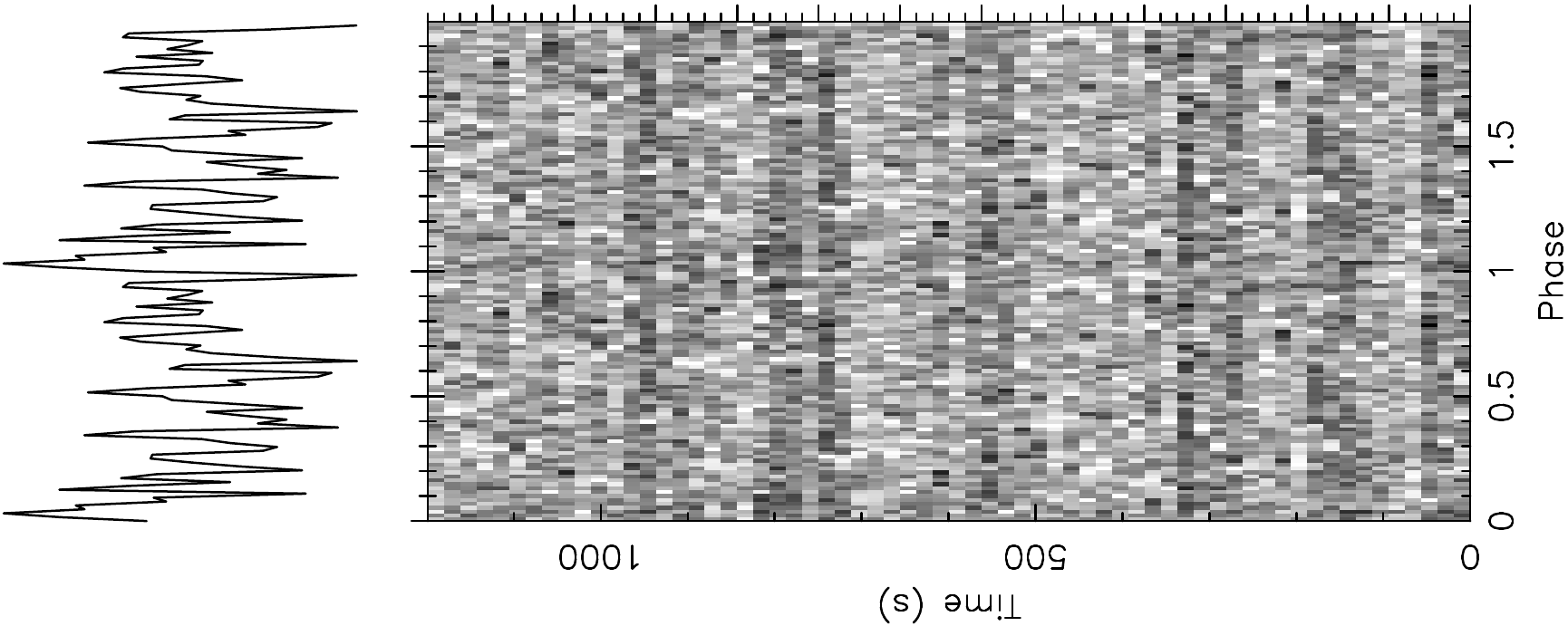}
  \caption{Same as Fig.\,\ref{fig:msps} but now for the 20-min
    discovery observation of PSR\,J1552+54. The left panel shows the
    result of coherently dedispersing the data at a
    $\mathrm{DM}=22.5$\,pc\,cm$^{-3}$, and then using incoherent
    dedispersion at the DM of the pulsar
    ($\mathrm{DM}=22.901$\,pc\,cm$^{-3}$). The right panel shows the
    same pulse profile when using incoherent dedispersion only. The
    pulse profile is no longer significant due to dispersive smearing
    within each of the 24-kHz channels.}
  \label{fig:j1552}
\end{figure}

Each of the coherently dedispersed filterbanks was then incoherently
dedispersed at steps of 0.002\,pc\,cm$^{-3}$ ranging from $-0.5$ to
0.5\,pc\,cm$^{-3}$ around the coherently dedispersed DM of the
filterbank data set. Here, we used the GPU accelerated brute force
dedispersion algorithm from the
\textsc{dedisp}\footnote{\url{http://code.google.com/archive/p/dedisp/}}
library by \citet{bbbf12} to generate incoherently dedispersed
time-series. Further processing was performed with programs from the
\textsc{presto} suite of pulsar search
software\footnote{\url{http://github.com/scottransom/presto}}
\citep{ran01}. In particular, power spectra for each incoherently
dedispersed time-series were created with \texttt{realfft}, and then
searched for Doppler-shifted periodic signals through a
frequency-domain acceleration search with a GPU-accelerated version of
\textsc{presto}'s
\texttt{accelsearch}\footnote{\url{http://github.com/jintaoluo/presto_on_gpu}}
\citep{rem02}. Finally, pulsar candidates were folded and their
parameters (spin period, DM and acceleration) optimized with
\texttt{prepfold}.

\begin{figure}
  \includegraphics[angle=270,width=\columnwidth]{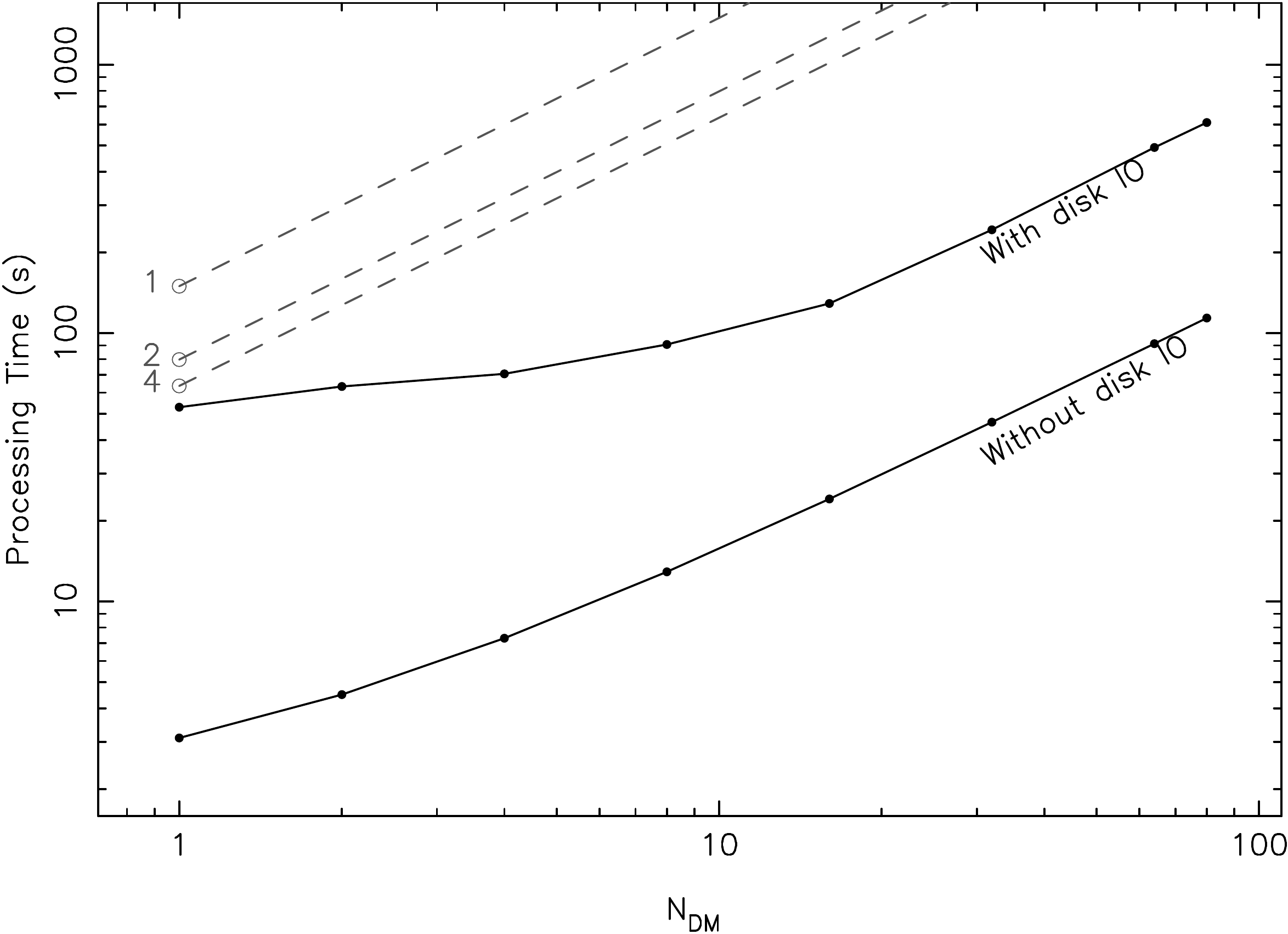}
  \caption{Performance of \texttt{cdmt} as a function of the number of
    coherently dedispersed DM trials. The input data consists of a
    20-min observation with 10 consecutive dual-polarization, complex
    valued, Nyquist sampled subbands of 195.31\,kHz bandwidth with
    frequencies between 114.93 and 116.88\,MHz. These are the
    lowest-frequency subbands from the LOFAR survey. An NVIDIA Titan X
    GPU was used with $N_\mathrm{bin}=65536$ samples,
    $n_\mathrm{d}=2048$ samples, $n_\mathrm{c}=8$ output channels per
    subband and $N_\mathrm{FFT}=100$ FFTs processed in parallel. The
    top curve shows the \texttt{cdmt} performance when reading in
    8.8\,GB of input data and writing 2.2\,GB of output data per DM
    trial. The bottom curve shows the \texttt{cdmt} performance when
    removing the disk reading and writing operations (\textsc{CUDA}
    memory copies, \textsc{cuFFT} operations and \textsc{CUDA} kernel
    operations). For comparison, the performance of \texttt{digifil}
    \citep{sb10} for single or multi-threaded operation (1, 2 or 4
    CPUs) on an 8-core Xeon E5-2630 CPU is shown with the gray open
    circles. Here, \texttt{digifil} computes a single DM trial. The
    extrapolated performance of \texttt{digifil} when coherently
    dedispersing multiple DM trials sequentially is shown with the
    dashed lines. Each point here is the average processing time of 10
    runs.}
  \label{fig:performance}
\end{figure}

The \texttt{cdmt} coherent dedispersion algorithm and the pulsar
search pipeline were tested on observations of pulsars J1810+1744 and
J2215+5135, which are binary radio millisecond pulsars visible at
radio frequencies of 150\,MHz \citep{kvh+16}.  Both pulsars were
blindly rediscovered through the pipeline described
above. Figure\,\ref{fig:msps} shows integrated pulse profiles as well
as pulse phase as a function of time for both millisecond pulsars. For
both pulsars coherent dedispersion is required, as the dispersion
smearing within a 24-kHz channel at a frequency of 135\,MHz
exceeds the pulsar spin period $P$; for PSR\,J1810+1744, the smearing
at $\mathrm{DM}=39.658$\,pc\,cm$^{-3}$ is $t_\mathrm{d}=3.3$\,ms,
while the spin period is $P=1.66$\,ms, while PSR\,J2215+5135 has
$P=2.61$\,ms with $t_\mathrm{d}=5.7$\,ms at
$\mathrm{DM}=69.194$\,pc\,cm$^{-3}$.

While preparing this manuscript, a new radio millisecond pulsar was
discovered as part of the LOFAR survey for \textit{Fermi} $\gamma$-ray
sources. It has $P=2.43$\,ms at a
$\mathrm{DM}=22.901$\,pc\,cm$^{-3}$. This discovery and the full
description of the survey will be published elsewhere (Pleunis et al.,
in prep.).

\section{Performance}\label{sec:performance}
With the dataset as described in \S\,\ref{sec:application} as input,
we can test the performance of \texttt{cdmt}. Here we specifically
focus on subbands with the lowest observing frequencies, where the DM
delays are largest. In particular, the input data consists of one part
with 10 consecutive subbands with frequencies between 114.93 and
116.88\,MHz. The performance of \texttt{cdmt} on a \textsc{Dragnet}
node using a single NVIDIA Titan X GPU is shown in
Fig.\,\ref{fig:performance}. Here, we use the following parameters;
$N_\mathrm{bin}=65536$ samples, $n_\mathrm{d}=2048$ samples,
$n_\mathrm{c}=8$ output channels per subband and $N_\mathrm{FFT}=100$
FFTs processed in parallel. Besides the processing time of all steps
in \texttt{cdmt}, we also show the processing time in the absence of
reading from and writing data to disk. Here the processing time is set
by the remaining steps (primarily the GPU operations; \textsc{CUDA}
memory copies, \textsc{cuFFT} operations and \textsc{CUDA} kernel
operations). As expected, we find that the GPU operations scale
linearly with the number $N_\mathrm{DM}$ of DM trials that are
coherently dedispersed, but that disk operations limit the
performance. The input data is 8.8\,GB while for each DM trial 2.2\,GB
of output needs to be written to disk.

Besides \texttt{cdmt}, coherently dedispersing a single DM is possible
with the \texttt{digifil} software. This program is part of the
\textsc{dspsr}\footnote{\url{http://dspsr.sourceforge.net}} software
package \citep{sb10} and performs coherent dedispersion on CPUs. In
Fig.\,\ref{fig:performance} we also provide the processing time of
\texttt{digifil} for the same input dataset on a Xeon E5-2630 CPU
using 1, 2 or 4 CPU threads. We find that for coherently dedispersing
a single dispersion measure on the \textsc{Dragnet} hardware, the
performance of \texttt{cdmt} is comparable to running \texttt{digifil}
with 4 threads. Like \texttt{cdmt}, the performance of
\texttt{digifil} is limited by disk operations, as using more than 4
threads does not decrease the processing time. When coherently
dedispersing multiple DM trials, \texttt{cdmt} will significantly
outperform \texttt{digifil} due to the data reuse in \texttt{cdmt}.
We do note that \texttt{digifil} optimizes the choice for FFT size
$N_\mathrm{bin}$ and data overlap $n_\mathrm{d}$ depending on the DM
and observing frequency of the actual input data. As a result,
\texttt{digifil} will be somewhat more efficient at low DMs and/or
higher observing frequencies compared to the fixed $N_\mathrm{bin}$
and $n_\mathrm{d}$ values used by \texttt{cdmt} that are chosen for
the lowest observing frequency of the input data and highest DM
required for the particular project.

The use of semi-coherent dedispersion offers higher time resolution at
all DMs, though requires having to perform more incoherent DM
trials. The semi-coherent dedispersion approach used for the LOFAR
\textit{Fermi} $\gamma$-ray source survey limits the dispersion
smearing to 0.07\,ms. Hence the time resolution and DM step size for
incoherent dedispersion can remain at 40.96\,$\upmu$s and
0.002\,pc\,cm$^{-3}$ over the full DM range. As a result, 40,000
incoherent DM trials are required. For comparison, using the
\textsc{PRESTO} dedispersion planning tool \texttt{DDplan.py}, a fully
incoherent search up to a maximum DM of 80\,pc\,cm$^{-3}$ would
require only 7,240 incoherent DM trials. However, without coherent
dedispersion the time and frequency resolution would have no
sensitivity for millisecond pulsars with periods below 10\,ms at DMs
above 10\,pc\,cm$^{-3}$ (see Fig.\,\ref{fig:ddplan}).

\begin{figure}
  \includegraphics[angle=270,width=\columnwidth]{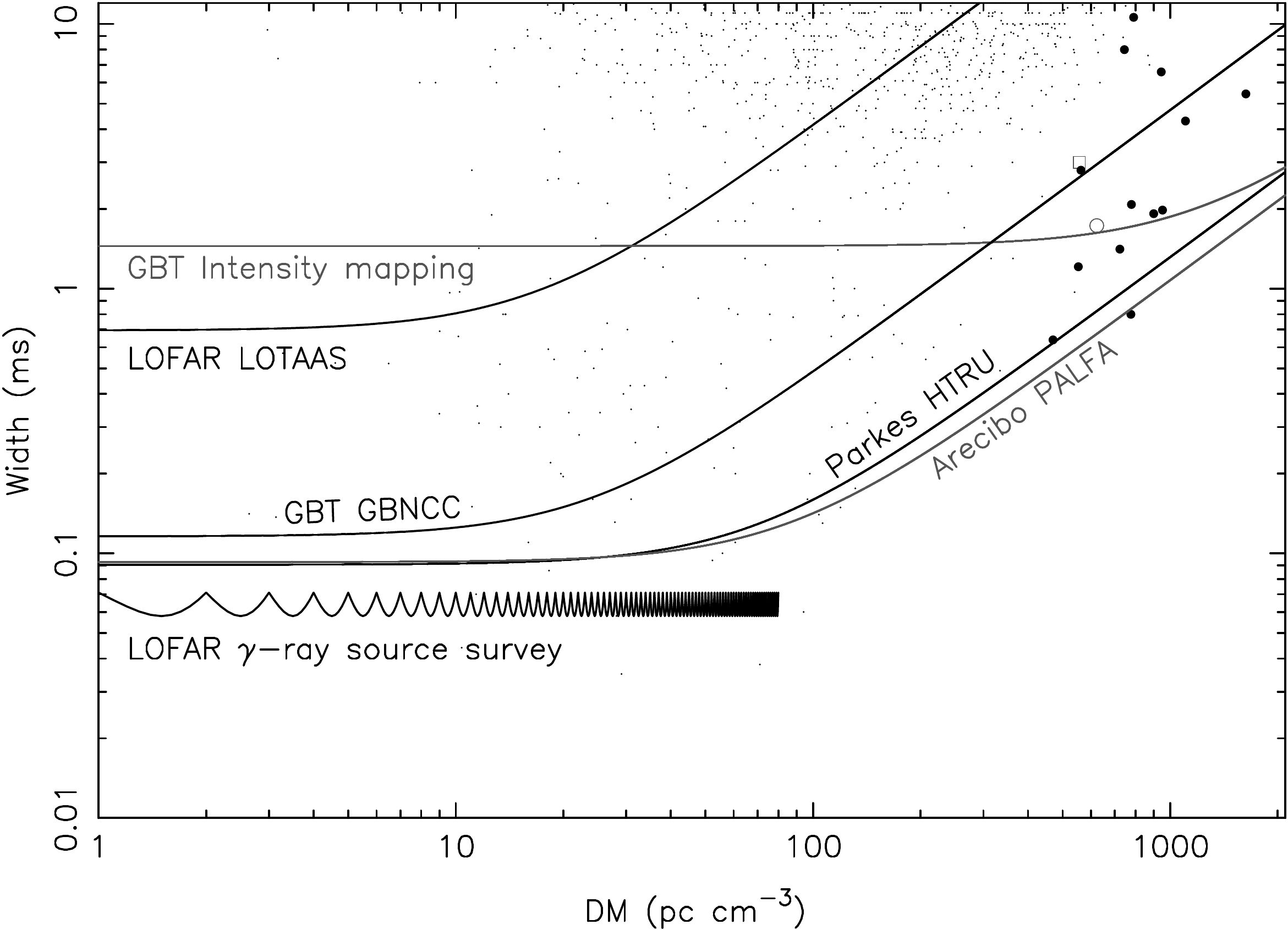}
  \caption{The DM and pulse width of pulsars (small dots) and Fast
    Radio Bursts (FRBs) discovered by the Parkes (large dots), Arecibo
    (open square) and GBT (open circle) radio telescopes. Pulsar
    parameters have been taken from the pulsar catalogue
    \citep{mhth05}, while FRB parameters originate from the FRB
    catalogue \citep{pbj+16}.  Pulse broadening curves for ongoing
    pulsar surveys with the Parkes (HTRU, 1.4\,GHz; \citealt{kjs+10}),
    Arecibo (PALFA, 1.4\,GHz; \citealt{lbh+15}), GBT (GBNCC, 350\,MHz;
    \citealt{slr+14}) and LOFAR (LOTAAS, 135\,MHz; \citealt{clh+14})
    radio telescopes are shown with solid lines. Also shown is the
    broadening curve for a Hydrogen intensity mapping experiment with
    the GBT (800\,MHz), which discovered an FRB (open circle,
    \citealt{mls+15}). The pulse broadening takes into account finite
    sampling times and dispersion smearing within channels, though
    neglects the effects of scattering. All surveys use incoherent
    dedispersion, except for the LOFAR \textit{Fermi} $\gamma$-ray
    source survey described in this paper. By combining coherent and
    incoherent dedispersion, this survey limits pulse broadening over
    the entire DM range that is being surveyed. }
  \label{fig:surveys}
\end{figure}

\section{Discussion}\label{sec:discussion}
Correcting for dispersion in surveys for pulsars and FRBs through
incoherent dedispersion is no longer a major computational
bottleneck. Recent implementations of incoherent dedispersion
algorithms on GPUs are fast enough to allow real-time processing
\citep{mks+11,bbbf12,slbn16}.  We have presented \texttt{cdmt}, the
next step in correcting for dispersion; an implementation to perform
coherent dedispersion for many different DM trials. These DM trials
serve as input for further incoherent dedispersion and allow
semi-coherent dedispersion searches for pulsars and FRBs. The
combination of coherent and incoherent dedispersion limits the
dispersion smearing and allows a more flexible choice of channel size
and sampling time at different observing frequencies and dispersion
measures.

We are using \texttt{cdmt} in an ongoing LOFAR survey of radio
millisecond pulsars associated with unidentified \textit{Fermi}
$\gamma$-ray sources. With this approach, we limit the dispersion
smearing at these low observing frequencies over the DM range being
surveyed, retaining sensitivity to short period pulsars at all DMs
being sampled (see Fig.\,\ref{fig:surveys}). The success of this
approach has already been demonstrated with the discovery of a new
millisecond pulsar (Fig.\,\ref{fig:j1552}).

Surveys for pulsars and fast transients are key science goals for
SKA1-Low and SKA1-Mid \citep{hpb+15,kbk+15,mkg+15}. Though the use of
semi-coherent dedispersion is currently not planned, the SKA1-Low
survey for pulsars at high Galactic latitudes can benefit from this
approach. As the SKA1-Low pulsar survey will likely operate at higher
frequencies (200 to 350\,MHz; \citealt{kbk+15}) in comparison to
LOFAR, dispersion smearing will be lower and hence fewer coherent DM
trials would be required. For an example setup of 100\,MHz of
bandwidth at a center frequency of 250\,MHz, a time resolution of
50\,$\upmu$s and 20\,kHz channels, coherently dedispersed DM step
sizes of 10\,pc\,cm$^{-3}$ would provide a maximum dispersion smearing
of 90\,$\upmu$s.

Semi-coherent dedispersion searches may also be useful at higher
observing frequencies. Recent FRB discoveries at observing frequencies
of 1.4\,GHz have DMs in excess of 1,000\,pc\,cm$^{-3}$
\citep{tsb+13,cpk+16,pbj+16} and for these the width of the pulses is
dominated by dispersion smearing within a channel
(Fig.\,\ref{fig:surveys}). Coherently dedispersing to a few DM trials
spaced at intervals of 10 to 100\,pc\,cm$^{-3}$ would keep the pulse
broadening due to dispersion smearing below 0.1\,ms and increase the
signal to noise of detected pulses, while also giving a more accurate
portrayal of the intrinsic pulse duration.

The \texttt{cdmt} software is continuing development, with new
features being planned. Features being worked on include implementing
the spectral kurtosis method \citep{ng10a,ng10b}, in order to reject
radio frequency interference when computing data offsets and scales to
remove its influence on the redigitization, and reading and writing
different input and output formats. The code is publicly available at
\url{http://github.com/cbassa/cdmt}.

\begin{small}
\section*{Acknowledgments}
We thank Vlad Kondratiev, Sotiris Sanidas and Alexander van Amesfoort
for their contributions to the design and construction of the
\textsc{Dragnet} computing cluster, as well as comments on drafts of
this manuscript.  The research leading to these results has received
funding from the European Research Council under the European Union's
Seventh Framework Programme (FP7/2007-2013) / ERC grant agreement
nr. 337062 (DRAGNET; PI Hessels).
\end{small}

\bibliographystyle{mnras}

\end{document}